\documentclass{INTERSPEECH2023}

\interspeechcameraready 

\usepackage{amsmath}
\usepackage{pifont}

\usepackage{algorithm}
\usepackage{algpseudocode}
\usepackage{tcolorbox}
\usepackage{caption}
\usepackage{lipsum}
\usepackage{mdwlist}
\usepackage{graphicx} 
\usepackage{verbatim}
\usepackage{verbatimbox}
\usepackage{fancyvrb}
\usepackage{wrapfig}
\usepackage{xcolor}
\usepackage{MnSymbol}
\usepackage{multirow}
\usepackage{bold-extra}
\usepackage{caption}
\usepackage{subcaption}
\usepackage{diagbox}
\usepackage{xspace}
\usepackage{mathtools}
\title{SlothSpeech: Denial-of-service Attack Against Speech Recognition Models}
\name{Mirazul Haque*$^1$, Rutvij Shah*$^1$, Simin Chen$^1$, Berrak Sisman$^1$, Cong Liu$^2$, Wei Yang$^1$}

\address{
  $^1$University of Texas at Dallas, USA\\
  $^2$University of California, Riverside, USA 
  }
\email{mirazul.haque@utdallas.edu, rutvij.shah@utdallas.edu, simin.chen@utdallas.edu, berrak.sisman@utdallas.edu, congl@ucr.edu, wei.yang@utdallas.edu}

\begin{document}

\maketitle
\begin{abstract}
Deep Learning (DL) models have been popular nowadays to execute different speech-related tasks, including automatic speech recognition (ASR). As ASR is being used in different real-time scenarios, it is important that the ASR model remains efficient against minor perturbations to the input. Hence, evaluating efficiency robustness of the ASR model is the need of the hour. We show that popular ASR models like Speech2Text model and Whisper model have dynamic computation based on different inputs, causing dynamic efficiency. In this work, we propose SlothSpeech, a denial-of-service attack against ASR models, which exploits the dynamic behavior of the model. SlothSpeech uses the probability distribution of the output text tokens to generate perturbations to the audio such that the efficiency of the ASR model is decreased.  We find that SlothSpeech-generated inputs can increase the latency up to 40X times the latency induced by benign input.



\end{abstract}

\section{Introduction}

Deep Learning (DL) models have been popular for executing different tasks like object recognition, machine translation, sentence classification, etc, with high accuracy. With the increasing popularity, DL models are also being used in speech-related tasks like Automatic Speech Recognition (ASR).
ASR is a task in which a given audio signal is transcribed to text. ASR has been significantly instrumental in tasks like caption generation of audio or using virtual speech assistants. 

Because of their usage in real-time scenarios, ASR models are needed to have high-efficiency robustness. Efficiency robustness evaluates if minor perturbation to the input would significantly decrease the efficiency of a model or not. However, unlike accuracy robustness 
~\cite{carlini2017towards,bojchevski2019adversarial,goodfellow2014explaining,zugner2018adversarial}, efficiency robustness has not been evaluated on ASR models. 
To evaluate efficiency robustness, we first need to investigate if any computation in the model is dynamic w.r.t input in nature, as it will cause dynamic efficiency based on different inputs.

First, we investigated the architectures of popular ASR models in Huggingface and found that there are mainly two types of decoder used in ASR models. We refer to the first type as static decoder, where the decoder generates a static number of word or character tokens and then removes unessential tokens. Popular ASR models like CTC models~\cite{zenkel2017comparison} use the static decoder. The second type is referred as dynamic decoder, where the number of generated tokens is dynamic based on input. Popular ASR models like Speech2Text~\cite{wang2020fairseq,DBLP:journals/corr/abs-2104-06678} and Whisper~\cite{radford2022robust} use dynamic decoder-based mechanisms.
As the dynamic decoder shows dynamic efficiency, the efficiency robustness of these  models is needed to be evaluated. In this work, we focus on the evaluation of the efficiency robustness of these dynamic decoder-based ASR models.

The efficiency robustness of a model ensures that the efficiency of the model is not impacted significantly by adding perturbation to the input. If a model lacks efficiency robustness, this could lead to fatalities. For example, an ASR system is used in an autonomous vehicle to recognize the driver's instructions. If the model is not efficiency-robust, the model could take a significant amount of time to respond, denying the service and leading to an accident. Hence, the efficiency robustness in a system is needed to be evaluated.


To evaluate the efficiency robustness of ASR models, first, we need to establish to the relation between input and dynamic efficiency. As mentioned earlier, for the dynamic decoder, number of generated tokens 
is dynamic w.r.t input and the number of generated tokens is dependent on a specific end-of-sentence($<EOS>$) token. If the occurrence of $<EOS>$ is delayed, the number of generated tokens is increased. The occurrence of any token is dependent on output probability distribution of the decoder. Hence, we need to modify the input, which can modify the output probability distribution of the decoder.

Based on the aforementioned relation between input and efficiency, we propose SlothSpeech, a denial-of-service attack on the ASR models. Denial-of-service attack increases the latency or the response time of the model significantly, ultimately denying the model service to the users. SlothSpeech is an iterative white-box attack that uses output probability distribution of different tokens to increase latency.

We evaluate SlothSpeech\footnote{*Both authors contributed equally to this research.} on three popular models: Speech2Text, Whisper, and Speech2Text2 ~\cite{wang2020fairseq,DBLP:journals/corr/abs-2104-06678,radford2022robust} and three popular datasets: LibriSpeech, OpenSLR and VCTK \cite{panayotov2015librispeech,korvas_2014,veaux2017cstr}. We evaluate SlothSpeech on two criteria: effectiveness and quality. Effectiveness measures the effect of SlothSpeech on the latency of the model. While quality measures the distance between adversarial input and benign input. We find that SlothSpeech-generated inputs can increase the latency up to 4000\% of the latency induced by benign input.

Our work makes the following contributions:

\begin{itemize}

    \item\textbf {Problem Characterization.} Our work is the first work to characterize the latency surge vulnerability in ASR models. 
    \item\textbf {Approach.} We propose a novel loss function to synthesize an iterative white-box attack. 

    \item\textbf {Experimentation.} We evaluate SlothSpeech on two criteria with three popular datasets and three popular models.

\end{itemize}

The rest of the paper is organized as follows: In Section \ref{sec:back}, we introduce the the background of dynamic-decoder-based ASR model and adversarial attack. In Section \ref{sec:ss}, we formulate the optimization problem of the denial-of-service attack on ASR models and we explain SlothSpeech approach. In Section \ref{sec:eval}, we discuss the evaluation results.

\section{Background}

\label{sec:back}
\subsection{Automatic Speech Recognition Systems}

\begin{figure}
    \centering
    \includegraphics[width=0.48\textwidth]{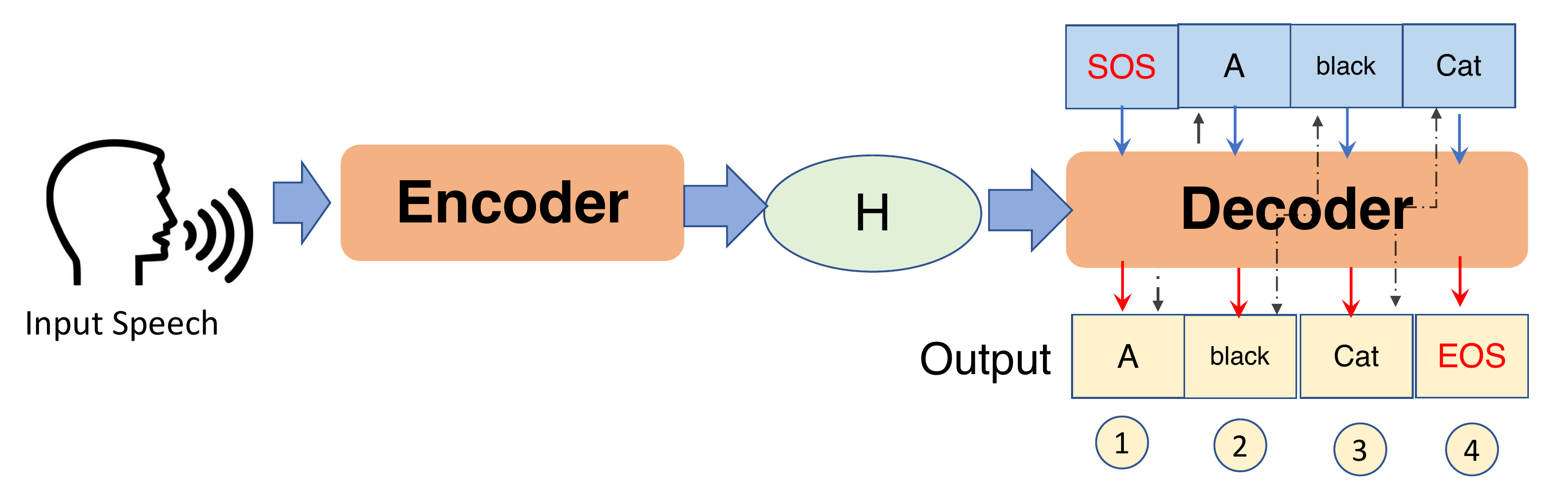}
    \caption{Working mechanism of dynamic-decoder-based ASR}
    \label{fig:asr}
\end{figure}
Given the input speech $\boldsymbol{x}$, the ASR systems compute the output probability for a sequence of tokens $\text{Pr}(\boldsymbol{y} | \boldsymbol{x})$ through the Bayes' theorem. 
\begin{equation}
\label{eq:bayes}
    \text{Pr}(\boldsymbol{y} | \boldsymbol{x}) = \prod_{u = 1} \text{Pr}(\boldsymbol{y_i)}  |  \boldsymbol{x}, \boldsymbol{y_1}, \cdots, \boldsymbol{y_{u - 1}})
\end{equation}
The computation process is shown in Equation \ref{eq:bayes}, where $\boldsymbol{y_i}$ is the $i^{th}$ output token. 
In this paper, our focus is on ASR systems that are based on the dynamic decoder architecture. Such systems comprise two key components - an encoder and a decoder. As illustrated in Figure \ref{fig:asr}, the encoder neural network is responsible for encoding the input speech into a hidden representation, while the decoder begins with a special token $SOS$ and generates subsequent output tokens iteratively by leveraging the decoder neural networks. The decoding process will continue until it reaches another token EOS.
A notable observation regarding the working mechanism of dynamic-decoder-based ASR systems is that the decoder is invoked more frequently for inputs that arrive at the end of the sequence (EOS). This observation implies that the ASR system tends to allocate more computational resources to inputs that have longer output sequences, thereby prioritizing the decoding of such sequences. Therefore, longer outputs lead to wastage of computational resources on the part of the victim ASR system.


\subsection{Adversarial Attacks against DNNs}
Recently, several adversarial attacks \cite{carlini2017towards,bojchevski2019adversarial,goodfellow2014explaining,zugner2018adversarial,haque2022corrgan} have been developed for targeting DNN-based systems. These attacks can create human-imperceptible adversarial perturbations, which, when applied to benign inputs, generate adversarial examples that can easily evade even the most advanced DNNs.
Based on the availability of the DNNs parameters, the adversarial attacks can be categorized as white-box attacks and black-box attacks.
Apart from correctness-based adversarial attacks, researchers have recently proposed denial-of-service attacks \cite{MirazILFO,haque2022ereba,chen2022deepperform,chen2022nicgslowdown,chen2022nmtsloth,haque2021nodeattack,chen2023dark} for targeting neural networks with dynamic decision routes. The objective of these attacks is to maximize the computational cost of the victim model, thereby decreasing its availability.

\section{SlothSpeech}
\begin{figure*}[t]
    \centering
    
    \includegraphics[width=0.8\textwidth]{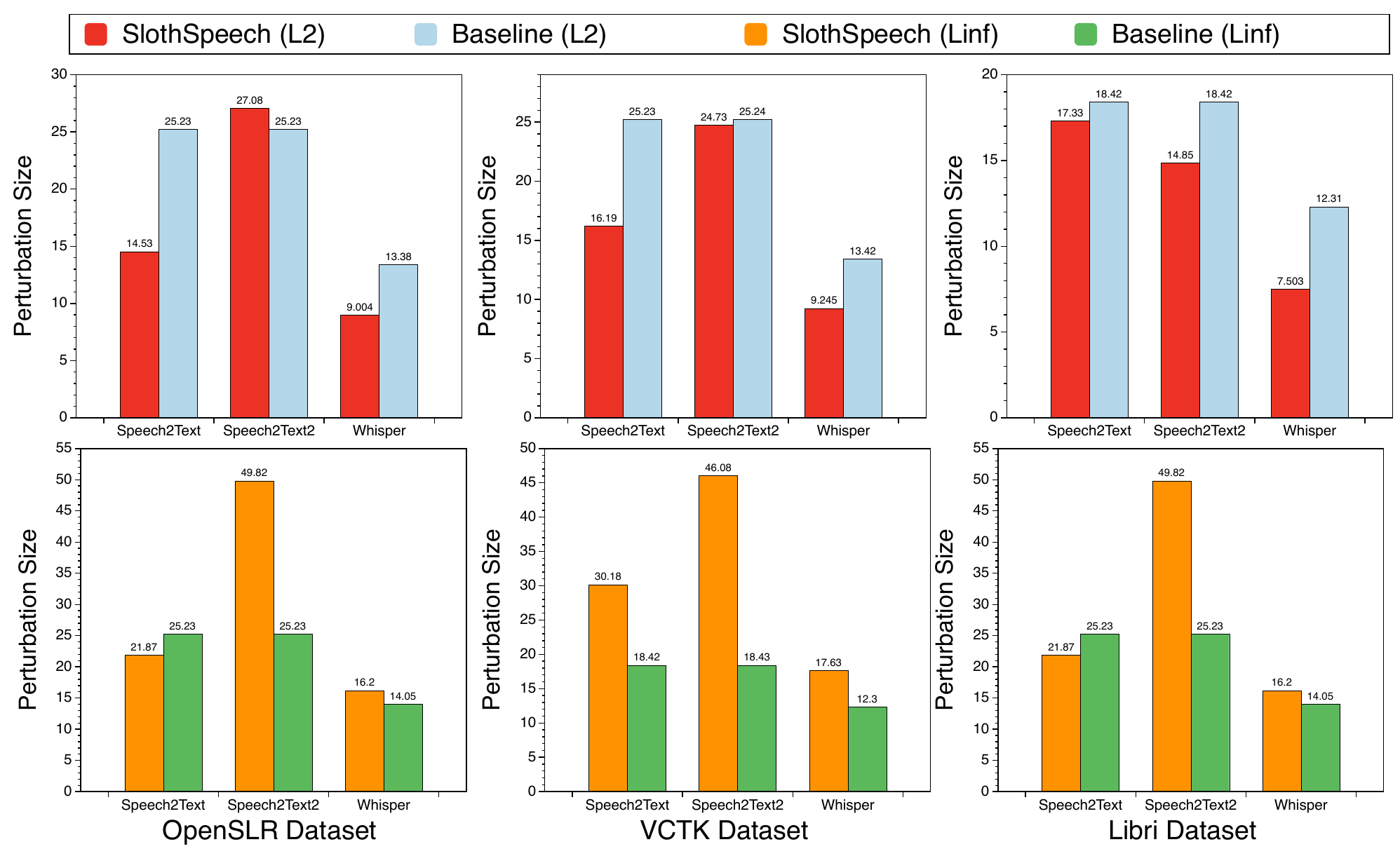}

    \caption{Comparison of Distance between SlothSpeech and Gaussian Noise.}
    \label{fig:quality}
\end{figure*}
In this section, we discuss the proposed approach SlothSpeech. First, we formulate the problem, then we focus on how we create the objective function, and finally, we explain the iterative optimization approach.
\label{sec:ss}
\subsection{Problem Formulation}

\begin{figure}
    \centering
    \includegraphics[width=0.48\textwidth]{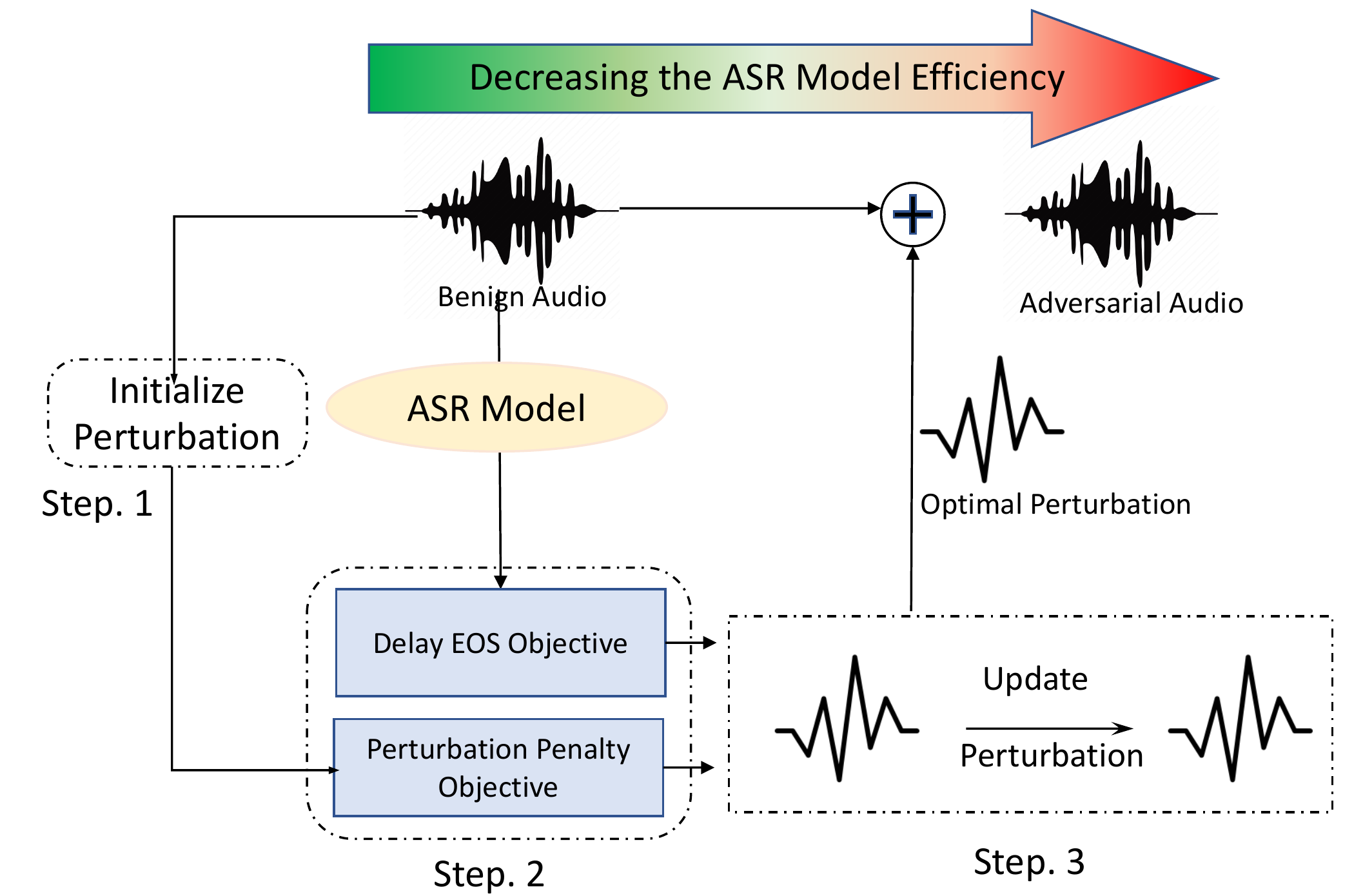}
    \caption{Design Overview of SlothSpeech}
    \label{fig:overview}
\end{figure}

Our objective is to produce audio that is imperceptible to humans and can decrease the efficiency of the victim's ASR model during inference. By reducing the efficiency of the ASR model, the adversary can deplete its computational resources, such as battery consumption, and make it unavailable, ultimately achieving the goal of denial-of-service. 
Our objective consists of two main factors: (i) reducing the efficiency of the victim's ASR model, and (ii) ensuring imperceptibility to humans.
We formulate our objective as an optimization problem, as shown in Equation \ref{eq:formulation},
\begin{equation}
    \label{eq:formulation}
        \Delta = \text{argmin}_{\delta} \, \text{Efficiency}_{f}(x + \delta) \qquad 
        s.t.  ||\delta|| \le \epsilon 
\end{equation}
where $x$ is an audio input fed to ASR system $f(\cdot)$, our aim is to generate an audio perturbation $\delta$ that minimizes the efficiency of the ASR system while also satisfying imperceptibility constraints."




\subsection{Differentiable Proxy of Latency}

As the $\text{Efficiency}_{f}$ is not a differentiable term w.r.t input, we need 
to find a proxy of $\text{Efficiency}_{f}$, which is differentiable. 
As discussed in the section \ref{sec:back}, the $\text{Efficiency}_{f}$ is dependent on the length of the output text. Also, the length of the output text is dependent on the occurrence of end-of-sentence or $<EOS>$ token. Our objective is to delay the occurrence of the $<EOS>$ token. To achieve this, we will first discuss how a token is selected for the output.

Formally, the ASR model's output is a sequence of probability distributions,  $f(x) = [p_1(x), p_2(x),\cdots, p_n(x)]$ and the output token sequences are $[t_1(x), t_2(x), \cdots, t_n(x)]$. Here, $t_i(x) =\text{argmax} (p_i(x))$.
Also, the likelihood of the output tokens and the EOS tokens are represented as $[p_1(x)^{t_1}, p_2(x)^{t_2}, \cdots, p_n(x)^{t_n}]$ and $[p_1(x)^{EOS}, p_2(x)^{EOS}, \cdots, p_n(x)^{EOS}]$ respectively.

To delay the occurrence of the EOS tokens, one approach would be to minimize the likelihood of the EOS tokens. But this approach would be resource-consuming because cross-entropy would be used on the large vocabulary. Instead of that, we convert the probability distribution $p_i$ for the multi-classification task to a binary classification task \textit{i.e.,},  is or not EOS token. The new probability distribution is represented as $q_i = [l_i(x)^{EOS}, \sum_j \l_i(x)^j - l_i(x)^{EOS}]$, where $j \neq EOS$. 

Our first objective is to decrease the likelihood of all EOS tokens in the output; hence, $\text{Efficiency}_{f}$ can be replaced by, 
$\mathcal{L}_{EOS}=  \frac{1}{n}\sum_{i=1}^{n}l_i(x+\delta)^{EOS}$
. However, considering only EOS tokens during the loss function can reduce the effectiveness of the attack. For the last token of the output sequence, EOS token's likelihood would be highest, which we need to decrease. Also, if we increase the likelihood of another specific token (token with second highest likelihood), it would increase the effectiveness of the attack because the specific token's likelihood would increase, and EOS token's likelihood would decrease. Hence the proxy of $\text{Efficiency}_{f}$ will be

$$\mathcal{L}_{EOS}=  \Big(\frac{1}{n}\sum_{i=1}^{n}l_i(x+\delta)^{EOS}+q_n(x+\delta)\Big)$$

where, $q_n$ represents the probability of the token with second highest likelihood in the last ($n^{th}$) token. If $P$ is the distance norm, and $c$ is a weight value defined by attacker, then the final optimization loss function can be represented by,

\begin{equation}
\label{eq:eos_loss}
    \mathcal{L} =||\delta||_P+c \cdot \mathcal{L}_{EOS}
\end{equation}

\subsection{Approach}
Algorithm \ref{alg:L2} and Figure \ref{fig:overview} show the details of SlothSpeech.
The SlothSpeech approach can be divided into three parts. (i) \textit{Initialize.} First, we initialize different variables that would be needed for synthesizing perturbation. (ii) \textit{Calculating loss function.} Next, we calculate loss function based on equation \ref{eq:eos_loss}. (iii) \textit{Update adversarial input.} Based on the optimization of loss function, we update the adversarial input. Below, we explain each step.

\noindent\textbf{Intitialize.} First, we initialize $\delta$ (Perturbation), $x^{*}$ (Final Adversarial Input) and $max_N$ (Maximum Number of Tokens), $iter$ (Iteration Index) in Line 2-3.. These values will be updated iteratively based on the optimization procedure.

\noindent\textbf{Calculating loss function.} In this step, we calculate the loss function that will be optimized. Based on equation \ref{eq:eos_loss}, we have two components to the loss function. First loss component, $\mathcal{L}_{EOS}$ is calculated based on the likelihood of different tokens (Line 6-7), whereas the second loss component $\mathcal{L}_{d}$ (Line 8) signifies the distance between adversarial and benign examples (\textit{e.g.}, $L_2$ and $L_\infty$). Then both components are added based on the weight $c$ (Line 9).

\noindent\textbf{Update adversarial input.} Finally, we update the adversarial input $hat{x}$ (Line 10), and check if the output length generated by $\hat{x}$ is better than the saved $max_N$ (Line 11-12). If yes,  the final adversarial input $x^{*}$ and $max_N$ is updated.

\begin{algorithm}[H]\small
        \flushright{
            \begin{algorithmic}[1]
\caption{SlothSpeech}\label{alg:L2}
\textbf{Input:} Benign input $x$, victim ASR $f(\cdot)$, maximum iteration number $T$, weight val $c$. \\
\noindent\textbf{Output:} Latency-based adversarial example $x^{*}$
  \State Initialize $\delta,x^{*},max_N$
   \State $iter \gets 0$
  \While{$iter \less T$}
        \State $\hat{x} =  x+\delta$
        \State $H =  f(\hat{x})$
        \State $\mathcal{L}_{EOS} \gets GetEOSLoss(H) $
        \State $\mathcal{L}_{d} \gets Distance(\hat{x},x)$
        \State$\mathcal{L}_{adv} \gets \mathcal{L}_{EOS}+c \cdot \mathcal{L}_{d}$ 
        \State $\hat{x} \gets \hat{x} + \frac{\partial \mathcal{L}_{adv}}{\partial \hat{x}}$

        \State $N \gets \text{GetLength}(\hat{x})$
        \State $max_N$, $x^*$ = Update($max_N$, $N$, $\hat{x}$)
    \EndWhile
    \State Return $x^{*}$

\end{algorithmic}
        }
    \end{algorithm}

\section{Evaluation}
\label{sec:eval}
We evaluate SlothSpeech based on two criterion: effectiveness and quality. 
\begin{table*}[hbtp]
  \centering
  \caption{\textbf{Mean and Max latency and number of outut tokens of the models
against SlothSpeech and baseline techniques. seed represents the original input, whereas abs and inc represent the absolute value and increase in metric respectively. All the experiments are performed on LibriSpeech, OpenSLR and VCTK datasets and three aforementioned Models.}}
\label{tab:Per_c10_2}
  \resizebox{\textwidth}{!}{
    \begin{tabular}{c|cl|rrrr|rrr|rrrr}
    \toprule
    \multirow{2}[2]{*}{Metric} & \multirow{2}[2]{*}{Dataset} & \multicolumn{1}{c|}{\multirow{2}[2]{*}{Model}} & \multicolumn{4}{c|}{Mean Absolute Value (Abs)} & \multicolumn{3}{c|}{Mean \% Increase (Inc)} &\multicolumn{4}{c}{Max Absolute Value (Abs)}  \\
          &       &       & \multicolumn{1}{c}{seed.} & \multicolumn{1}{c}{Gaussian.} & \multicolumn{1}{c}{$L_2$ (Ours)} & \multicolumn{1}{c|}{$L_\infty$ (Ours)} & \multicolumn{1}{c}{Gaussian.} & \multicolumn{1}{c}{$L_2$ (Ours)} & \multicolumn{1}{c|}{$L_\infty$ (Ours)} & \multicolumn{1}{c}{seed.} & \multicolumn{1}{c}{Gaussian.} & \multicolumn{1}{c}{$L_2$ (Ours)} & \multicolumn{1}{c}{$L_\infty$ (Ours)} \\
    \midrule
    \multirow{4}[1]{*}{\# of Tokens} & \multirow{3}[1]{*}{Libri} & Speech2Text &     22.20  &   22.17    &   \textbf{1001.00}      & \textbf{1001.00}      & 0     &  \textbf{4415.00}     &  \textbf{4415.00}  &85.00 & 80.00 & \textbf{1001.00} &\textbf{1001.00} \\
          &       & Whisper & 22.95       &     22.84  &   \textbf{42.82}    &   42.58   &    0.00   &    \textbf{86.50}   & 85.50& 98.00 &96.00&151.00&\textbf{173.00}\\
          &       & Speech2Text2 & 24.36       & 29.46      &   152.50    &  \textbf{429.61}   &    20.90  &    526.00   & \textbf{1663.00}&56.00 &463.00 &\textbf{1001.00} &\textbf{1001.00} \\
          \cline{2-14}
          & \multirow{3}[0]{*}{OpenSLR} & Speech2Text &  91.05     &   102.99    &  \textbf{1001.00}     &   \textbf{1001.00}    &   13.11    &  \textbf{999.00}     & \textbf{999.00} & \textbf{1001.00} &\textbf{1001.00} &\textbf{1001.00} &\textbf{1001.00} \\
          &       & Whisper &  21.65    &  19.98     &   36.85   &    \textbf{38.36}  &   -7.70\%    &   70.20    & \textbf{77.18} & 48.00 &58.00 &249.00 &\textbf{366.00}\\
          &       & Speech2Text2 &  24.32     &  12.33     &   \textbf{315.62}     &  292.53     & -49.00      &   \textbf{1197.00}    & 1102.00 &157.00 & 22.00 &\textbf{501.00} &\textbf{501.00} \\
          \cline{2-14}
          & \multirow{3}[0]{*}{VCTK} & Speech2Text &  22.35     &   32.59    &  \textbf{1001.00}     &   1000.00    &   46.00    &  \textbf{4378.00}     & 4374.00 & 199.00 &\textbf{1001.00} &\textbf{1001.00} &\textbf{1001.00}\\
          &       & Whisper &  17.66     &  17.68     &   32.03   &    \textbf{32.59}  &   0.00    &   81.30    & \textbf{84.50} &50.00 &43.00&102.00&\textbf{441.00}\\
          &       & Speech2Text2 &  18.05     &  12.12     &   \textbf{260.50}     &  206.57     & -32.80      &   \textbf{1343.00}    & 1044.00&54.00 &19.00 &\textbf{501.00}&\textbf{501.00} \\
          \midrule
    \multirow{4}[1]{*}{Latency (ms)} & \multirow{2}[0]{*}{Libri} & Speech2Text &            218.00  &   214.53    &    8722.00      &\textbf{8746.00}     &    -1.80   &  3900.00     & \textbf{3911.00} &875.00 &823.00 & 9039.00 &\textbf{9224.00} \\
          &       & Whisper & 376.00      &     369.00 &   \textbf{637.00}    &  630.00    &  -1.80     &   \textbf{69.00}    &  67.50&1329.00 &1348.00 &2068.00 &\textbf{2293.00}      \\
          &       & Speech2Text2 & 567.00      &  1738.00   &  10722.00    & \textbf{10725.00}    &  206.00     &   1791.00   &  \textbf{1791.00}      & 2237.00 &60493.00 & \textbf{99793.00} &70287.00\\
          \cline{2-14}
          & \multirow{2}[1]{*}{OpenSLR} & Speech2Text & 932.00     &   905.00    &   8767.00      &  \textbf{8778.00}     &    -2.80   &   840.60    &  \textbf{842.00} & 8778.00 &8660.00 &9036.00 &\textbf{9038.00}\\
          &       & Whisper & 376.00      & 352.00      &   600.00   &   \textbf{618.00}    &  -6.30     & 59.50      & \textbf{64.30} & 759.00 &902.00 & 3587.00 &\textbf{5640.00}\\
          &       & Speech2Text2 & 893.00      & 558.00      &   \textbf{16101.00}   &     13565.00  &  -37.40     & \textbf{1703.00}      & 1428.00 & 8996.00 & 944.00 & 32190.00 &\textbf{36124.00} \\
          \cline{2-14}
          & \multirow{3}[0]{*}{VCTK} & Speech2Text &  242.00     &   364.00    &  \textbf{10091.00}     &   9960.00    &   50.40    &  \textbf{4069.00}    & 4015.00 & 2003.00 &10219.00 &\textbf{10996.00}&10835.00 \\
          &       & Whisper &  308.00 & 304.00    &   499.00   &    \textbf{506.00}  &   -1.20    &   62.00   & \textbf{64.20} & 764.00 &681.00& 1430.00 &\textbf{6006.00}\\
          &       & Speech2Text2 &  720.00     &  502.00     &   \textbf{12704.00}     &  9604.00     & -30.77      &   \textbf{1670.00}    & 1233.00 &11269.00 &906.00 & \textbf{35377.00} &29965.00 \\
          
    \bottomrule
    \end{tabular}%
    }
  \label{tab:addlabel}%
\end{table*}%

\subsection{Experimental Setup}
\label{subsec:setup}


\noindent\textbf{Datasets and Models.} For evaluation, LibriSpeech dataset~\cite{panayotov2015librispeech}, OpenSLR~\cite{korvas_2014}, and VCTK dataset~\cite{veaux2017cstr} have been used for synthesizing the adversarial examples. 
We use three popular ASR models: Speech2Text~\cite{wang2020fairseq}, Whisper~\cite{radford2022robust} and Speech2Text2~\cite{zenkel2017comparison}. All the pre-trained weights are gathered from Huggingface.

\noindent\textbf{Baseline and Metric} As this is the first denial-of-service attack on ASR models, we use Gaussian noise as baseline.
We examine two metrics to reflect the ASR models' computational costs in order to measure the effectiveness of SlothSpeech in increasing the victim ASR models' computational costs. The first is the number of tokens generated by the decoder (hardware-independent metric), and the second is the latency of ASR models (hardware-dependent metric) in handling an input.
 We first measure the absolute computational costs (Abs.) of the benign inputs and then generate adversarial examples. After we compute the computational cost increments (Inc.). 
 For quality evaluation, we use the distance between adversarial and benign input as metric, while for transferability evaluation, we measure percentage increase in latency as metric.

 \noindent\textbf{Implementation Details.} 
 For this work, we use $c$ value as 1 and $T$ value as 100 to generate adversarial inputs. We set $max\_length$ of the tokens as 1001, and as distant norm, we use $L_2$ and $L_\infty$ norms. Only for Speech2Text2 model tested with OpenSLR and VCTK datasets, we use 500 as $max\_length$, because extending that limit caused significant load on the GPU.


\subsection{Effectiveness} As mentioned earlier, we measure the effectiveness of SlothSpeech by measuring computational latency and number of tokens in the output. Table \ref{tab:Per_c10_2} shows our results of the effectiveness of SlothSpeech and the Gaussian noise.  
We show both the mean absolute values of the metric in the table and the mean percentage increase in the metric due to perturbation. (seed being the original input). Also, we show maximum absolute values for number of tokens and latency achieved by seed input and different technique-generated inputs.

It can be noticed that all three models are vulnerable against SlothSpeech and, SlothSpeech-generated inputs perform significantly better than baseline w.r.t increasing computation in ASR models For Speech2Text model, all the adversarial examples induce maximum length of the output token. The results reflect that Speech2Text model is the least efficiency-robust against SlothSpeech. Speech2Text2 model also shows low robustness against SlothSpeech, however the percentage increase in latency for Speech2Text is higher than Speech2Text2. Whisper model has shown higher robustness against SlothSpeech than the other two models. Inputs generated using both distant norms have similar effectiveness, however, for S2T-Libri pair, inputs generated through $L_\infty$ norm have significantly higher effectiveness than $L_2$ norm.

\begin{center}
\begin{tcolorbox}[colback=gray!10,
                  colframe=black,
                  width=8cm,
                  arc=1mm, auto outer arc,
                  boxrule=0.9pt,
                 ]
 \textbf{Summary.} SlothSpeech generated input outperforms the baseline by significantly increasing the latency of the ASR models (up to 4000\%).
\end{tcolorbox}
\end{center}
\subsection{Quality} 

We evaluate quality of adversarial examples w.r.t magnitude of the perturbation added to the audio signal due to SlothSpeech and baseline. Figure \ref{fig:quality} shows the results. We use bargraph to show
the mean of different perturbations added to the input by SlothSpeech and baseline. It can be observed that mean value of L2 perturbation for baseline and SlothSpeech is similar for all case scenarios, however, for $L_\infty$ norm, SlothSpeech perturbation is slightly higher than the baseline. Although except S2T2 model, the mean perturbation of SlothSpeech is always similar to or lower than baseline for both norm. Hence, it can be noted that with similar mean perturbation, SlothSpeech has a significantly higher effectiveness than baseline.

\begin{center}
\begin{tcolorbox}[colback=gray!10,
                  colframe=black,
                  width=8cm,
                  arc=1mm, auto outer arc,
                  boxrule=0.9pt,
                 ]
 \textbf{Summary.} The mean perturbation magnitude added  to SlothSpeech generated input is similar to the magnitude of the Gaussian noise.
\end{tcolorbox}
\end{center}


\section{Conclusion}

In this work, we propose SlothSpeech\footnote{https://github.com/0xrutvij/SlothSpeech}, a white-box denial-of-service attack that can decrease the efficiency of the ASR models significantly. SlothSpeech uses the likelihood of output tokens to generate adversarial inputs. We evaluate SlothSpeech on three popular datasets and three popular models. We find that SlothSpeech-generated inputs can increase the model latency up to 40 times more than the benign input. However, in this work, we do not focus on improving the efficiency robustness of the ASR models.

\clearpage
 

\bibliographystyle{IEEEtran}
\bibliography{mybib}

\end{document}